\documentclass[10pt,epsfig]{article}
\newcommand\beq{\begin{equation}}
\newcommand\eeq{\end{equation}}
\newcommand\bea{\begin{eqnarray}}
\newcommand\eea{\end{eqnarray}}

\usepackage{amsmath,graphicx,latexsym,eufrak,mathrsfs,bbm,pxfonts,parskip}
\begin{document}
\vspace{-2.0cm}
\bigskip

\baselineskip=15pt
\centerline{\Large \bf SU(N) Coherent States and Irreducible Schwinger Bosons} 
\vskip .8 true cm

\begin{center} 
{\bf Manu Mathur}\footnote{manu@bose.res.in} and
{\bf Indrakshi Raychowdhury} \footnote{indrakshi@bose.res.in}
\vskip 0.2 true cm

S. N. Bose National Centre for Basic Sciences \\ 
JD Block, Sector III, Salt Lake City, Calcutta 700098, India

\end{center} 
\bigskip

\centerline{\bf Abstract}

\noindent We exploit the SU(N) irreducible Schwinger boson to construct 
SU(N) coherent states. This construction of SU(N) coherent state is 
analogous to the construction of the simplest Heisenberg-Weyl coherent states.  
The coherent states belonging to irreducible representations of SU(N) are labeled 
by the eigenvalues of the $(N-1)$ SU(N) Casimir operators and are characterized  by $(N-1)$ complex 
orthonormal vectors describing the SU(N) group manifold.

%\noindent PACS: ~02.20.-a 

\section{\bf Introduction}

The concept of coherent states was introduced by Schr\"odinger \cite{sc} in the context of a 
harmonic oscillator. These harmonic oscillator coherent states, also called canonical coherent states,
have been widely used in physics \cite{klauder,glauber,sudarshan,zhang,perg}. The next important coherent
states are spin coherent states or SU(2) coherent states which are associated with angular
momentum or the SU(2) group. Like canonical coherent states, they too have found wide
applications in different branches of physics such as quantum optics, statistical mechanics,
nuclear physics and condensed matter physics \cite{klauder,zhang,perg,auer,marshal}. It is known that 
these spin coherent states can also be constructed using harmonic oscillators by exploiting either the 
Holstein–Primakov or the Schwinger boson representation of the SU(2) Lie algebra \cite{sch,rad,erik}. This harmonic
oscillator formulation of spin coherent states is appealing because of its simplicity as this construction 
is analogous to the simplest and oldest canonical coherent state construction. Further, unlike the standard construction of 
coherent states \cite{perg}, this method does not require any knowledge of  group representations or 
group elements and their actions on a particular weight vector, leading to many technical simplifications 
(see section \ref{su22}). Infact, these SU(2) coherent states in terms of harmonic oscillators have been 
implicitly  contained in the seminal work of Schwinger \cite{sch} way back in 1952. 
The aim of the present work is to show that the above simple, uniform and explicit construction of  canonical 
and spin or SU(2) coherent states can also be easily extended to all higher SU(N) groups. This is in 
contrast to the standard construction of  SU(N) coherent states (i.e., by applying the SU(N) group elements on 
a particular weight vector) which is known to become more and more tedious as N increases \cite{gitman,nemoto}. In 
the past, this problem has led to various different approaches \cite{gitman,nemoto,md,mm,pur,mu,gnu} to construct 
SU(3) and SU(N) coherent states.   
In \cite{gitman}  SU(N) coherent states belonging to SU(N) symmetric representations are constructed by using 
fixed order polynomials  of complex N-plets. In \cite{nemoto} a very special characterization of SU(N) 
group elements is exploited to construct SU(N) coherent states. In \cite{md,mm,pur} Schwinger boson representation 
of SU(N) Lie algebra is used to construct SU(N) coherent states. 
In \cite{mu} the Schwinger oscillator representation of SU(3) coherent states is analyzed to discuss 
its relationship with the standard harmonic oscillator coherent states.  In \cite{gnu} SU(3) coherent states 
are constructed using a special parametrization of SU(3) group elements. 
%In \cite{pu} SU(N,M) coherent states 
%in the symmetric representations are constructed using the corresponding Schwinger bosons. 
%Motivated by the resulting simpliﬁcations, we recently
%generalized this harmonic oscillator formulation of coherent states to SU(N) group [7]. In
%this work, we further exploit the above ideas to construct SU (2) and SU (3) charge coherent
%states deﬁned on S 3 and S 5 , respectively. The coherent states carrying SU (2) and SU (3)
%(non-Abelian) charges in two- and three-mode Fock spaces have been discussed in the past
%[5, 8–10]. However, they are deﬁned on full complex planes and are different from the SU (2)
%and SU (3) charge coherent states discussed in this paper which are deﬁned on the compact

The present work exploits the SU(N) irreducible Schwinger  bosons \cite{su3isb,sunisb} to construct SU(N) 
coherent states belonging to an arbitrary irreducible representation of SU(N). By definition, 
the SU(N) irreducible Schwinger bosons creation operators carry all the symmetries of SU(N) irreducible 
representations (see section \ref{su3} and \ref{sun}). As a result all SU(N) irreducible representation 
states are  monomials (not polynomials) 
of SU(N) irreducible Schwinger boson creation operators  acting on the corresponding vacuum states.  
Therefore, as in the 
case of canonical and spin coherent states, they are the natural candidates for constructing the SU(N) 
coherent states. These coherent states are defined on the SU(N) group manifold  
characterized by an orthonormal set of 
$(N-1)$  complex SU(N) vectors. In addition, the SU(N) coherent states are  labeled by the integer 
eigenvalues of $(N-1)$  Casimir operators which are the $(N-1)$ types of Schwinger boson number operators 
(see sections \ref{su3} and \ref{sun}).    

\noindent The organization of the paper is as follows. We start with a very brief discussion on
harmonic oscillator or canonical coherent states because  the SU(N) coherent state construction in this 
work is analogous to this simplest construction. In section \ref{su22}, we illustrate this similarity 
by constructing SU(2) coherent states in terms of a doublet of harmonic oscillators or equivalently SU(2) 
Schwinger bosons. As mentioned earlier, this construction has been exploited by 
Schwinger \cite{sch} to compute SU(2) recoupling coefficients. At the end of section \ref{su22},    
the simplifications obtained by this Schwinger boson approach to coherent states over the 
standard approach of applying a SU(2) group element on a particular weight vector are highlighted. 
In sections \ref{su3} and \ref{sun} we further extend  the above SU(2) construction to SU(3) and SU(N) 
respectively. These SU(3) and then SU(N) extensions of SU(2) coherent states are again trivial as they 
correspond to: 
\begin{itemize} 
\item  replacing SU(2) Schwinger boson doublet by $(N-1)$ SU(N) irreducible Schwinger bosons N-plets, 
\item  replacing SU(2) group manifold (i.e., a doublet of complex numbers) by SU(N) group manifold (i.e., 
$N-1$ N-plets of complex numbers). 
\end{itemize} 
%This amounts to replacing SU(2) doublet of complex numbers which describes  SU(2) group manifold
%by $(N-1)$ mutually orthonormal complex N-plets which describe SU(N) group manifold. 
The section \ref{su3} on SU(3) is added to make the transition from SU(2) (section \ref{su22}) to 
SU(N) (section \ref{sun}) easy.       

\noindent  In the simplest example of the Heisenberg-Weyl group, 
the Lie algebra contains three generators. It is defined in terms 
of creation annihilation operators $(a, a^{\dagger})$ satisfying
\beq
[a,a^{\dagger}]= {\cal I}, ~~~[a,{\cal I}] =0, ~~~[a^{\dagger}, {\cal I}]=0 ~.
\label{eqn1} 
\eeq
This algebra has only one infinite dimensional unitary irreducible 
representation.  The states within this representation  
are the  occupation number states 
$\vert n\rangle \equiv 
{(a^{\dagger})^n \over \sqrt {n!}} \vert 0\rangle$  
with $n=0, 1, 2 ...$ . 
The coherent states of the Heisenberg-Weyl group are defined over a complex  manifold as: 
\bea\label{wcs}
\vert z\rangle_{[\infty]} ~=~ \exp (z a^{\dagger} ) ~\vert 0\rangle =\sum_{n=0}^{\infty} F_n (z) ~\vert n\rangle. 
\eea
In (\ref{wcs}) the subscript $[\infty]$ on the coherent states is the irreducible representation index. 
It implies that these coherent states are defined over the infinite dimensional irreducible representation of 
the group.  The sum in (\ref{wcs}) runs over all the basis vectors $|n\rangle$ belonging to this 
infinite dimensional representation.  The coefficients:  
\beq
F_n (z) ~=~ {z^n \over \sqrt{n!}} 
\label{hwsf} 
\eeq
are the coherent state expansion coefficients which are analytic functions of the group manifold coordinate 
z. The resolution of identity property of the coherent state (\ref{wcs}) follows from the group transformation 
property. Let us define the operator: ${\cal O}_{[\infty]} \equiv \int  e^{-|z|^2} dz d\bar{z} 
%exp{-z\bar{z}}
~|z\rangle_{[\infty]} ~{}_{[\infty]}\langle z|$.  
%with K a constant and ${\cal I}$ is the identity operator 
Under the Heisenberg Weyl group element \cite{perg} $g_{{\textrm hw}} \equiv exp \left(i~\alpha+w a 
- \bar{w}a^\dagger\right): |z\rangle_{[\infty]} \rightarrow  e^{i~\alpha + zw -{w\bar{w}\over 2}} 
|z- \bar{w}\rangle_{[\infty]}$. It is trivial to see that the operator ${\cal O}_{[\infty]}$ defined above is 
invariant under $g_{{\textrm hw}}$. Therefore, by Schurs lemma it is proportional to unity operator.  

The purpose of this work is to generalize 
(\ref{wcs}), (\ref{hwsf}) for Heisenberg-Weyl group to SU(N) for arbitrary N (see (\ref{cs2}), (\ref{su2sf}) 
for SU(2);  (\ref{cs333}), (\ref{su3sf}) for SU(3) and (\ref{cs33}), (\ref{sfsun}) for SU(N)). 
We start with SU(2) construction \cite{sch,rad} first.   
%The main feature of this construction is that, an 
%expansion of the coherent states in terms of basis vectors of a given 
%representation with analytic functions of complex variables ($F_n (z)$) as 
%coefficients is obtained.  It is easy to see that Eq. (\ref{wcs}) provides a resolution of identity 
%with the measure $d\mu(z) = dzd{\bar{z}}$. \\

\section{\bf $SU(2)$ Coherent States}
\label{su22}

\begin{figure}[t]
\begin{center}
\includegraphics[width=0.4\textwidth,height=0.13\textwidth]
{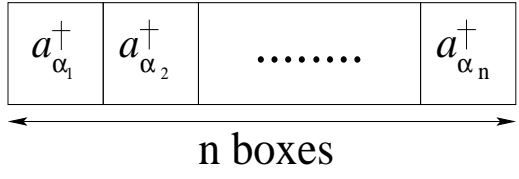}
\end{center}
\caption{SU(2) Young table in the $n=2j$ representation. The monomial state (\ref{su2irrep}) carries the horizontal 
permutation symmetries of this Young tableau.}   
\label{su2yt}
\end{figure}

The Heisenberg Weyl coherent state construction can be readily generalized to the simplest compact group 
SU(2) by utilizing the Schwinger representation  of SU(2) Lie algebra:
$[J^{\mathrm a},J^{\mathrm b}] = i \epsilon^{\mathrm abc} J^{\mathrm c}$. 
We define \cite{sch}: 
\bea
J^{\mathrm a} ~\equiv ~\frac{1}{2} ~a^{\dagger}_\alpha ~(\sigma^{\mathrm a} )^{\alpha}{}_{\beta} ~a^\beta.
\label{sch} 
\eea
In (\ref{sch}) $\sigma^{\mathrm a}$ with $\mathrm a= 1,2,3$ denote the three Pauli matrices. 
The doublet of harmonic oscillator creation and annihilation operators 
${a^\alpha} $ and ${a}^{\dagger}_\alpha$ or equivalently Schwinger bosons in (\ref{sch}) satisfy 
the simple bosonic commutation relations $[a^\alpha,a^{\dagger}_\beta] = \delta^{\alpha}{}_{\beta}$ 
with $\alpha,\beta =1,2$.  
The vacuum state $\vert 0,0\rangle$ of these two oscillators will be  denoted by $\vert 0 \rangle$. 
Under SU(2) transformations the Schwinger boson creation operators transform as doublets: 
\bea 
a^\dagger_\alpha \rightarrow a^\dagger_\beta ~ \left(\exp{i \theta^{\mathrm a} 
{\sigma^{\mathrm a}\over 2}} \right)^\beta_{~~\alpha}. 
%~~~~   a^\alpha \rightarrow \left(exp -i \theta^{\mathrm a} {\sigma^{\mathrm a}\over 2}\right)^\alpha_\beta a^\beta.
\label{su2gt}
\eea    
%We  note that $
%[Q^{\mathrm a},a^{\dagger}_{\alpha}] = a^{\dagger}_\beta {1 \over 2}({\sigma^{\mathrm a}})^{\beta}{}_{\alpha} 
%$.
%Hence  $(a^{\dagger}_{1}, 
%a^{\dagger}_{2})$ transform like a SU(2) doublet. This fundamental 
%representation of SU(2) will be  denoted by the Young diagram given by a single row of boxes. 
%Therefore, we can realize 
%all SU(2) irreducible representation on the Hilbert space of 
%harmonic oscillators created by creation operators acting on the vacuum 
%which is a direct product of the vacuum states for $a_{1}$ and $a_{2}$. 
The defining equations (\ref{sch}) imply that the SU(2) Casimir operator is simply 
the total number operator: 
\bea  
{\cal C} \equiv  \sum_{\alpha=1}^{2} {a}^{\dagger}_{\alpha}{a}^{\alpha} 
\equiv {a}^{\dagger}\cdot{a}. 
\label{su2cas} 
\eea 
The eigenvalues of ${\cal C}$ will be denoted by $n$. The various states in the irreducible representation 
$n (=2j)$ are:  
%SU(2), characterized by the eigenvalues of the Casimir in (\ref{su2cas}) as $n$,  can also be defined by its Young diagram which 
%is obtained by arranging $n$ number of boxes in a row.  This particular SU(2) irreducible representation 
%in terms of SU(2) Schwinger bosons is 
%given by: 
\bea
\vert\alpha_{1}\alpha_{2}....\alpha_{n}\rangle_{[n]} \equiv a^{\dagger}_{\alpha_{1}}a^{\dagger}_{\alpha_{2}}
.....a^{\dagger}_{\alpha_{n}} |0\rangle  
\label{su2irrep} 
\eea 
The corresponding SU(2) Young tableau is shown in Figure (\ref{su2yt}). Note that the state in (\ref{su2irrep}) 
is invariant under all $n!$ permutations of the SU(2) indices $\alpha_1,\alpha_2,\cdots ,\alpha_n$. This is 
because all SU(2) creation operators on the right hand side of (\ref{su2irrep}) commute amongst themselves. 
In other words, the SU(2) Schwinger boson creation operators carry the the symmetries of the SU(2) Young 
tableau\footnote{This obvious symmetry argument  will not be true for higher SU(N) (sections \ref{su3} and \ref{sun})  
leading to the definition of SU(N) irreducible Schwinger bosons. In terms of SU(N) irreducible Schwinger bosons the 
SU(N) irreducible states will be monomials like (\ref{su2irrep}).}  which is shown in Figure (\ref{su2yt}). 
Therefore, the $(n+1)$ states in (\ref{su2irrep}) 
belong to SU(2) irreducible representation with total angular momentum $j = {n\over 2}$. 
%so that each $a^\dagger_\alpha$ creates a single box in the SU(2) Young tableau and also maintains all of its symmetry properties.
%With the harmonic oscillator
%creation and annihilation operators, $SU(2)$ coherent 
%states can be  obtained by directly generalizing (\ref{wcs}). 

\noindent The SU(2) group manifold $S^3$ can also be described by a doublet of of complex numbers $(z^1, z^2)$ 
of unit magnitude:
\bea
|z|^{2} \equiv |z_1|^2 + |z_2|^2 =1. 
\label{su2cons} 
\eea
This is because  any SU(2) matrix ${\mathcal U_2}$ can be written as: 
\bea
\mathcal U_2  = \left( \begin{array}{cc} 
z_1 & z_2 \\
-z_2^* & z_1  \\ 
\end{array} \right)
\label{su2m}
\eea
with $\mathcal U_2^\dagger \mathcal U_2 = \mathcal U_2 \mathcal U_2^\dagger = 1, ~|\mathcal U_2| =1$. 
At this stage one can trivially combine the SU(2) irreducible states in (\ref{su2irrep}) 
and the SU(2) group manifold coordinates in (\ref{su2cons}) to construct the generating function 
of the SU(2) coherent states: 
\bea
\vert z\rangle \equiv  \vert z^1, z^2\rangle =  exp \left(z \cdot a^{\dagger} \right)  |0\rangle 
%\nonumber \\ 
 =  \sum_{n=0}^{\infty} \frac{\left({z} \cdot {a}^\dagger \right)^n}{n!} ~\vert 0\rangle 
%\nonumber \\
= \sum_{n=0}^{\infty} |z\rangle_n.  
%a^{\dagger}_{\alpha_{1}}a^{\dagger}_{\alpha_{2}}..  a^{\dagger}_{\alpha_{n}}|0\rangle \nonumber \\
%&=& F^{\alpha_{1}\alpha_{2}...\alpha_{n}}
%\vert {\alpha_{1}}{\alpha_{2}}..  {\alpha_{n}}\rangle  
\label{cs2} 
\eea
% $|z_1,z_2\rangle_{n=2j}$ in the various SU(2) irreducible representations characterized by $n=2j$: 
\noindent Above 
%In (\ref{cs2}), 
$z \cdot a^\dagger \equiv z^1a^\dagger_1 +z^2 a^\dagger_2$ and 
 $|z\rangle_{[n]}$ is the coherent state in the SU(2) representation $j=n/2$:  
\bea 
|z\rangle_{[n]} = 
\sum_{\alpha_1,\alpha_2,\cdots,\alpha_n=1}^{2}  F^{\alpha_{1}\alpha_{2}...\alpha_{n}}(z)  
{a^\dagger_{\alpha_{1}}a^\dagger_{\alpha_{2}} \cdots a^\dagger_{\alpha_{n}}
|0\rangle}
=\sum_{\alpha_1,\alpha_2,\cdots,\alpha_n=1}^{2}  F^{\alpha_{1}\alpha_{2}...\alpha_{n}}(z)  
\underbrace{|\alpha_{1}{\alpha_{2} \cdots \alpha_{n=2j} \rangle}_{[j]}}_{{\textrm SU(2)~irrep.}~ j= \frac{n}{2}}   
\label{su2cs2} 
\eea 
Like in Heisenberg Weyl case (\ref{hwsf}), the SU(2) coherent state structure functions in the irreducible 
representation $j ={n \over 2}$ are:  
\bea 
F^{\alpha_{1}\alpha_{2}...\alpha_{n}}(z^1,z^2)  \equiv  \frac{1}{n!} z^{\alpha_{1}} z^{\alpha_{2}}... z^{\alpha_{n}}. 
\label{su2sf} 
\eea
Note that they are  analytic functions of group manifold coordinates. 
%Like in the case of Heisenberg Weyl coherent states (\ref{hwsf}) they are the structure functions 
%of  the SU(2) coherent states in the $j = n/2$ representation. 
The resolution of identity property again follows from the group transformation laws. 
The coherent state structure in (\ref{cs2}) and the the SU(2) transformations (\ref{su2gt}) imply 
that under group transformations:~$|z^1,z^2\rangle_{[n]} \rightarrow |z^{\prime 1},z^{\prime 2} 
\rangle_{[n]}$ where $\left(z^{\prime 1},z^{\prime 2}\right)$ are the SU(2) rotated coherent state co-ordinates: 
\bea 
{z^{\prime}}^{\alpha} =\left( \exp i (\theta^{\mathrm a} \frac{\sigma^{\mathrm a}}{2})
\right)^{\alpha}{}_{\beta} ~z^{\beta}.
\label{uut}
\eea
\noindent
Therefore, under the SU(2)transformations the coherent states $|z\rangle \equiv |z_{1},z_{2}\rangle$ 
transform amongst themselves on $S^3$ as the 
constraint (\ref{su2cons}) remains invariant under (\ref{uut}). 
Again we define the operator: 
%analogous to (\ref{xxx}) Resolution of identity for SU(2) coherent state is obtained as: 
\beq
{\cal O}_{[n]} \equiv \int d\mu(z) \left(|z\rangle_{[n]}~{}_{[n]}~\langle z|\right) = 
\int ~d^{2}z^{1}d^{2}z^{2} \,\delta (~|z^1|^2 + |z^2|^2-1~)~|z \rangle _{[n]}~ {}_{[n]} \langle z |. 
%~=~ K~ \sum_{{\alpha_{1}},....  {\alpha_{n}}=1}^{2}
%\vert a^{\dagger}_{\alpha_{1}}a^{\dagger}_{\alpha_{2}}..
%a^{\dagger}_{\alpha_{n}}\rangle \langle a^{\dagger}_{\alpha_{1}}a^{\dagger}_{\alpha_{2}}..
%a^{\dagger}_{\alpha_{n}}\vert
%~=~ K\mathcal I
\eeq
%\beq
%\int ~d^{2}z^{1}d^{2}z^{2}\, |z \rangle _C~
%{}_{C} \langle z | ~=~ K ~ \sum_{{\alpha_{1}},....
%{\alpha_{C}}=1}^{2}
%\Vert a^{\dagger}_{\alpha_{1}}a^{\dagger}_{\alpha_{2}}..
%a^{\dagger}_{\alpha_{C}}\rangle \langle a^{\dagger}_{\alpha_{1}}a^{\dagger}_{\alpha_{2}}..
%a^{\dagger}_{\alpha_{C}}\Vert
%~=~ K \mathcal I
%\eeq
The operator  ${\cal O}_{[n]}$ is invariant under all  SU(2) transformations of the coherent states 
$|z \rangle _{[n]}$. Therefore, 
\bea
\left[Q^{\mathrm a},{\cal{O}}_{[n]}\right] = 0, ~~~~ \forall {\mathrm a} =1,2,....,8
\eea 
\noindent The  Schur's Lemma implies that ${\cal{O}}_{[n]}$ is proportional to identity operator. 
Before generalizing (\ref{cs2}) 
to SU(N), it is illustrative to briefly mention the standard group theoretical coherent state 
construction procedure \cite{perg}.  We characterize the SU(2) group elements U by the
Euler angles, i.e, $U(\theta,\phi,\psi) \equiv  exp -i \phi J_3  exp -i 
\theta J_2 exp -i \psi J_3$.   The SU(2) coherent states are constructed as: 
%The standard group theoretical definition
%(\ref{gp}) takes $\vert 0>_{j}$ in (\ref{gp}) to be the highest
%weight state $\vert j, j>$ and is of the form:
\bea
\vert \theta,\phi, \psi \rangle_{j} ~=~ U(\theta,\phi,\psi) ~\vert j, j\rangle
 ~,
=~ \sum_{m=-j}^{+j} C_{m}(\theta,\phi,\psi)  ~\vert j, m\rangle ~, \nonumber 
%\label{st}
\eea
\noindent The coefficients $C_{m}(\theta,\phi,\psi)$ are given by,
$$C_{m}(\theta,\phi,\psi) = e^{-i (m \phi+j \psi)} \big[{ 2j! \over (j+m)!(j-m)!}\big]^{1
\over 2} \big[sin {\theta \over 2}\big]^{j-m} \big[cos {\theta \over 2}
\big]^{j+m}.$$  It is clear that the corresponding construction is difficult for higher SU(N) group 
as we need to know all the SU(N) representations, Euler angles and the group elements to implement 
this procedure. On the other hand,  the  coherent states in (\ref{cs2}) are straightforward generalization 
of the Heisenberg-Weyl coherent states  in (\ref{wcs}) and bypass all the problems mentioned above. 
Our aim in this work is to further extend this simple coherent state construction to SU(N) with arbitrary N. 
As we will see the only new input required for this purpose is the replacement of SU(2) Schwinger bosons 
by SU(N) irreducible Schwinger bosons \cite{su3isb, sunisb}. We first deal with SU(3) group in detail. 

\section{SU(3) Coherent States}
\label{su3} 

We  start with a  brief review of  SU(3) irreducible Schwinger bosons \cite{su3isb,sunisb} 
%which will  replace SU(2) Schwinger bosons in (\ref{su2irrep}) and (\ref{su2cs2}) to 
and then construct SU(3) coherent states. 

\subsection{\bf The Irreducible Schwinger Boson Representations of $SU(3)$ } 

\begin{figure}[h]
\begin{center}
\includegraphics[width=0.5\textwidth,height=0.25\textwidth]
{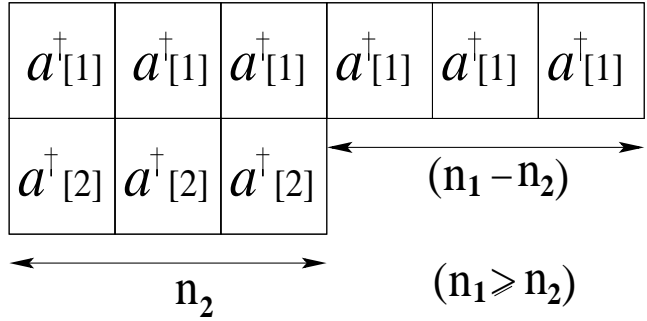}
\end{center}
\caption{SU(3) Young table in the $[n_1,n_2]$ representation. The monomial state (\ref{su3ir}) 
in terms of SU(3) irreducible Schwinger bosons and not (\ref{mssu3}) carries all the symmetries 
of this SU(3) Young tableau.}   
\label{su3yt2}
\end{figure}

As the rank of SU(3) group is two, we need two independent triplets to construct any arbitrary irrep of SU(3). 
We take them to be two independent harmonic oscillator triplets or equivalently 
Schwinger boson and denote them by: $a^{\dagger}_{\alpha}[1]$ and $a^{\dagger}_{\alpha}[2]$. 
The SU(3) generators in terms of these Schwinger bosons are:
\bea
\label{su3gen}
Q^{\mathrm a}= a^\dagger [1]\,\frac{\lambda^{\mathrm a}}{2}\,a[1] +a^\dagger [2]\,\frac{\lambda^{\mathrm a}}{2}\,a[2], ~~
a=1,2, \cdots ,8.  
\label{abc} 
\eea
In (\ref{abc})  $\lambda^{\mathrm a} $'s are the Gell-Mann matrices. 
%The $6$ Harmonic oscillators present in (\ref{su3gen}) creates a $6$ dimensional Hilbert space 
%$\mathcal H^{6}_{\mathrm {HO}}$.  
Under SU(3) transformations both types of Schwinger boson creation operators  transform as triplets: 
\bea 
a^\dagger_\alpha[i] \rightarrow a^\dagger_\beta[i]~ 
\left(\exp{i \theta^{\mathrm a} {\lambda^{\mathrm a}\over 2}} \right)^\beta_{~~\alpha} ~~~~~~~ i=1,2. 
%  a^\alpha[i] \rightarrow \left(exp -i \theta^{\mathrm a} 
%{\lambda^{\mathrm a}\over 2}\right)^\alpha_\beta a^\beta[i] ~~~~~~~~~i =1,2.
\label{su3gt}
\eea    
The  defining equations (\ref{su3gen}) immediately imply that the two SU(3) Casimirs commuting with all 
the generators are the two total number operators: 
\bea 
{\mathcal C}[1] \equiv {\cal N}_1 \equiv  a^{\dagger}[1] \cdot a[1], ~~~~  
{\mathcal C}_2 \equiv {\cal N}_2  \equiv  a^{\dagger}[2] \cdot a[2].  
\label{su3c} 
\eea 
It is obvious that $\left[{\cal C}[i],Q^a\right] = 0, \forall i=1,2; ~ a=1,2, \cdots ,8$  
as each $Q^a$ contains one creation and one annihilation operator of either type [i=1] or [i=2]. 
Their eigenvalues are  denoted by  $n_1$ and $n_2$ respectively. The corresponding 
irreducible representation with $n_1 \ge n_2$ is denoted\footnote{Note that $n_1$ and $n_2$ are the  numbers 
of two SU(3) triplets and not triplets and anti-triplets. As an example, $[1,1]$ represents 
the anti-triplet $3^*$ representation.} by $[n_1,n_2]$. 
The associated  SU(3) Young tableau is shown in Figure (\ref{su3yt2}). 

\noindent The monomial states 
constructed out of the two SU(3) fundamental Schwinger bosons:
%\footnote{We are using the following notations: 
%$\vert \cdots \rangle$ denote the SU(N) reducible states while $|\cdots\rangle_I$ 
%denote the SU(N) irreducible states.}: 
\bea 
\vert\alpha_{1}\alpha_{2} \cdots \alpha_{n_1}; \beta_1\beta_2 \cdots \beta_{n_2}\rangle 
\equiv 
\Big(a^{\dagger}_{\alpha_{1}}[1]a^{\dagger}_{\alpha_{2}}[1] .....a^{\dagger}_{\alpha_{n_1}}[1]\Big) 
\Big(a^{\dagger}_{\beta_{1}}[2]a^{\dagger}_{\beta_{2}}[2] .....a^{\dagger}_{\beta_{n_2}}[2]\Big) 
|0\rangle  
\label{mssu3} 
\eea
are eigenstates of ${\cal C}[1]$ and ${\cal C}[2]$ with eigenvalues $n_1$ and $n_2$ 
respectively.  However, unlike SU(2) case where the monomial  states (\ref{su2irrep}) are SU(2) 
irreducible, the corresponding monomial states (\ref{mssu3}) are SU(3) reducible and do not form the 
SU(3) irreducible representation $[n_1,n_2]$. 
%This is because the SU(3) Schwinger bosons do not carry all the symmetries  of SU(3) Young tableau. 
Like in SU(2) case, the monomial state (\ref{mssu3}) carries the horizontal symmetries of SU(3) Young 
tableaux as the Schwinger bosons creation operators on the right hand side commute amongst themselves. 
However, the vertical antisymmetry needs to be imposed to get SU(3) irreducibility. We achieve 
this by imposing the 
%In addition 
%we need to follow the following ordering to avoid multiplicity of irreps that $i^{th}$ row of an Young 
%tableaux is created by $a^{\dagger}[i]$.
%The reason for this is the existance of two non-trivial SU(3) invariant operators \cite{mosh} 
%$a^{\dagger}[1] \cdot a[2]$ and $a^{\dagger}[2] \cdot a[1]$ in addition to the two 
%trivial number operator in (\ref{su3c}) which are the Casimirs. 
%These imply that we can always 
%impose the 
following constraint \cite{sunisb,mosh}: 
\bea 
\hat{L}_{12} \equiv a^{\dagger}[1] \cdot a[2] =0 
\label{con3} 
\eea
on the monomial states (\ref{mssu3}). 
As an example, the simple  $[n_1=1,n_2=1]$ irreducible representation of SU(3) corresponds to the 
anti-triplet $3^*$ and the corresponding states are given by: 
\bea 
\vert \alpha;\beta\rangle_{[n_1=1,n_2=1]}  = - ~\vert \beta; \alpha \rangle_{[n_1=1,n_2=1]}  
\equiv 
\left(a^{\dagger}_{\alpha}[1] a^{\dagger}_\beta[2]- a^{\dagger}_{\beta}[1] a^{\dagger}_\alpha[2]\right) 
|0\rangle. 
\label{asr} 
\eea 
The three states in (\ref{asr}) trivially satisfy the constraint (\ref{con3}) as they are 
anti-symmetric. Therefore, they belong to SU(3) irreducible representation $3^*$.
%In terms of Young tableaues a general $[n_1,n_2]$ 
%SU(3) irreducible representation is represented by drawing  $n_1$ number of boxes in first row and 
%$n_2 $ number of boxes in second row with $n_1 \ge n_2$ as shown in figure \ref{su3yt2}. 
% Each of these  boxes in the first and second 
%row correspond to  $a^\dagger[1]$ and  $a^\dagger[2]$ respectively. 
%Note that unlike SU(2) the SU(3) Young tableaux or irrep 
%is not obtained by a simple monomial of $n_i$ numbers of $a^\dagger[i]$'s for $i=1,2$ as any Young tableaux 
%is symmetric along each row and antisymmetric along each column. Thus the Schwinger boson representation for 
%an arbitrary SU(3) irrep is indeed too complicated to do any practical calculation. Although the horizontal 
%symmetry of an Young tableaux is inbuilt in schwinger bosons itself the vertical antisymmetry has to 
%be implemented such that the final state satisfy the  constraint $a^\dagger[1]\cdot a[2]\approx 0$ . 
In  recent works \cite{sunisb,su3isb} we have defined SU(3) irreducible Schwinger bosons $A^\dagger[i]$ 
with  $i=1,2$ as: 
\bea
A^{\dagger}_\alpha[1]&=& a^{\dagger}_\alpha[1] \\
A^{\dagger}_\alpha[2] &=& a^{\dagger}_\alpha[2]- \frac{1}{\mathcal N_1-\mathcal N_2+2}\left(a^\dagger[2]\cdot a[1] 
\right) a^{\dagger}_\alpha[1]
\label{su3irsb} 
\eea
The SU(3) irreducible Schwinger bosons in (\ref{su3irsb}) are constructed such that \cite{su3isb,sunisb}: 
\bea
\left[ \left(a^{\dagger}[1] \cdot a[2]\right) ,A^{\dagger\alpha}[1] \right]\simeq 0 \nonumber \\
\left[ \left(a^{\dagger}[1] \cdot a[2]\right) ,A^{\dagger\beta}[2] \right]\simeq 0
\eea
Where, `$\simeq 0$' means that the commutators are weakly zero. In other words,  the above 
commutators annihilate all SU(3) irreducible states satisfying (\ref{con3}). 
%\bea 
%\left(a^{\dagger}[1] \cdot a[2]\right)  A^{\dagger\alpha}[1] ~\big|{\alpha_1 \alpha_2 \ldots\alpha_{n_1}; 
%\beta_1\beta_2 \ldots \beta_{n_2}} \big\rangle_{[n_1n_2]} = 0, \nonumber \\
%\left(a^{\dagger}[1] \cdot a[2]\right)  A^{\dagger\alpha}[2] ~\big|{\alpha_1 \alpha_2 \ldots\alpha_{n_1}; 
%\beta_1\beta_2 \ldots \beta_{n_2}} \big\rangle_{[n_1n_2]} = 0.
%\label{xyz} 
%\eea
%Above $| \cdots \rangle_{[n_1n_2]}$ denotes  SU(3) irreducible state already satisfying the constraint (\ref{con3}). 
The irreducible Schwinger bosons further satisfy: 
\bea 
\left[A^\dagger_\alpha[1], A^\dagger_\beta[1] \right] = 0, 
~~~~~~~~\left[A^\dagger_\alpha[2], A^\dagger_\beta[2] \right] = 0. 
%\left[A^\dagger_\alpha[1], A^\dagger_\beta[2] \right] = ????????.
\label{su3cr} 
\eea  
The defining equations (\ref{su3isb}) imply that their transformation properties are exactly same 
as (\ref{su3gt}): 
\bea 
A^\dagger_\alpha[i] \rightarrow A^\dagger_\beta[i]~ 
\left(\exp{i \theta^{\mathrm a} {\lambda^{\mathrm a}\over 2}} \right)^\beta_{~~\alpha} ~~~~~~~ i=1,2. 
%  a^\alpha[i] \rightarrow \left(exp -i \theta^{\mathrm a} 
%{\lambda^{\mathrm a}\over 2}\right)^\alpha_\beta a^\beta[i] ~~~~~~~~~i =1,2.
\label{su3gta}
\eea    
We now consider the most general monomial state constructed out of SU(3) irreducible Schwinger bosons: 
%As a result all SU(3) irreducible states are simple monomials in terms of SU(3) irreduble Schwinger bosons: 
\bea
\label{su3ir} 
\big\vert\alpha_1 \alpha_2 \ldots\alpha_{n_1};
\beta_1\beta_2 \ldots \beta_{n_2} \big\rangle_{[n_1n_2]}  \equiv 
\Big(A^{\dagger}_{\beta_1}[2] A^{\dagger}_{\beta_2}[2] 
\ldots A^{\dagger}_{\beta_{n_2}}[2]\Big) 
\big(A^{\dagger}_{\alpha_1}[1]  A^{\dagger}_{\alpha_2}[1] \ldots  
A^{\dagger}_{\alpha_{n_1}}[1]\Big)  
\big|0\big\rangle.  
\eea 
This monomial state directly creates the SU(3) Young tableau with $n_1$ and $n_2$ boxes in the first and second 
rows respectively. This is because the inbuilt constraint (\ref{con3}) ensures the vertical antisymmetry 
and  the commutators (\ref{su3cr}) ensure the horizontal symmetries of SU(3) Young tableau. 
Therefore, the monomial states in (\ref{su3ir}) belong to $[n_1,n_2]$ irreducible representation of SU(3). 
%As an example, we again consider the anti-triplet state in (\ref{asr}): 
%$$\vert \alpha;\beta\rangle_{[1,1]}  \equiv  
%A^\dagger_\beta[2] A^\dagger_\alpha[1] |0 \rangle = 
%\frac{1}{2} \left(a^\dagger_\alpha[1] a^\dagger_\beta[2] -a^\dagger_\beta[1] a^\dagger_\alpha[2] \right)|0\rangle.$$ 
%Similarly, the SU(3) octet state ($n_1=2,n_2=1$) is given by: 
%\bea 
%|\alpha_1,\alpha_2;\beta_1\rangle_{[2,1]}& =& A^\dagger_{\beta_1}[2] A^\dagger_{\alpha_2}[1] A^\dagger_{\alpha_1}[1] 
%|0 \rangle  \nonumber \\
%&=& \frac{1}{3} \left[\big(a^\dagger_{\alpha_1}[1]a^\dagger_\beta[2]-a^\dagger_{\beta}[1]a^\dagger_{\alpha_1}%[2]\big)
%a^\dagger_{\alpha_2}[1] + \big(\alpha_1 \leftrightarrow \alpha_2\big)\right] \nonumber  
% \left(a^\dagger_{\alpha_2}[1]a^\dagger_\beta[2]-a^\dagger_{\beta}[1]a^\dagger_{\alpha_2}[2]\right)
%a^\dagger_{\alpha_1}[1]\right]. \nonumber 
%\eea   
%Thus all SU(3) irreducible states are monomials of SU(3) irreducible Schwinger bosons.   
We now exploit this simple fact to further extend the definition of Heisenberg Weyl, SU(2) coherent states 
(\ref{wcs}) and (\ref{cs2}) to SU(3) group. 

\subsection{Construction of SU(3) Coherent States}

Similar to SU(2) case (\ref{su2cons}) and (\ref{su2m}), the eight dimensional SU(3) group manifold can be 
characterized by two complex triplets: $z_\alpha[1]$ and $z_\alpha[2]$  ($\alpha=1,2,3$) which 
satisfy the orthonormality constraints:
\bea
\label{zc3}
\bar z[1]\cdot z[1]=1=\bar z[2]\cdot z[2] ~~,~~ \bar z[1]\cdot z[2]=0.
\eea 
This is because any SU(3) matrix ${\mathcal U_3}$ can be written as:
\bea
\mathcal U_3 = \left( \begin{array}{ccc} 
z_1[1] & z_1[2]   & (\bar z[1]\wedge \bar z[2])_1  \\
z_2[1] & z_2[2] &  (\bar z[1]\wedge \bar z[2])_2  \\
z_3[1] & z_3[2]  &  (\bar z[1]\wedge \bar z[2])_3 
\end{array} \right)
\label{acc} 
\eea
with $\mathcal U_3 \mathcal U_3^\dagger =  \mathcal U_3^\dagger  \mathcal U_3  = 1$ and 
$det(\mathcal U_3) = |\mathcal U_3| =1$ due to the orthonormality constraints (\ref{zc3}).  

We define the SU(3) coherent states generating function as: 
\bea 
\big|z[1],z[2]\big\rangle \equiv \exp \left( z[2]\cdot  A^{\dagger}[2] \right)
\exp\left( z[1]\cdot A^{\dagger}[1]\right) \big|0\big\rangle 
\label{cs333}
\eea 
Note that this construction is SU(3) extension of SU(2) coherent state generating function  (\ref{cs2}). 
We can  project  SU(3)  coherent state in the representation $[n_1,n_2]$ by considering the corresponding 
term in the generating function (\ref{cs333}): 
%denoted by the  set of particular values of the SU(3) casimirs, 
%i.e., $A^{\dagger}[1]\cdot A[1] = \mathcal N_{1}~~\& ~~ 
%A^{\dagger}[2].A[2] =\mathcal N_{2}$ as $n_1$ and $n_2$ where, $n_1\ge n_2$ one gets,
\bea 
\label{cs23}
\big|z[1],z[2]\big\rangle_{[n_1,n_2]} &\equiv&  \frac{\left( z[2]\cdot  A^{\dagger}[2] \right)^{n_2}}{n_2!} 
\frac{\left( z[1]\cdot  A^{\dagger}[1]\right)^{n_1}}{n_1!} \big|0\big\rangle \nonumber  \\ 
&=& \sum_{\alpha_1..\alpha_{n_1}=1}^{3} \sum_{\beta_1..\beta_{n_2}=1}^{3} 
F^{\alpha_1..\alpha_{n_1};\beta_1..\beta_{n_2}}\big(z[1],z[2]\big) ~
\underbrace{\big|\alpha_1\alpha_2...\alpha_{n_1};\beta_1\beta_2...\beta_{n_2} \big\rangle_{[n_1n_2]}}_{
{\textrm SU(3)~irrep.}~[n_1,n_2]}.  
\eea 
In (\ref{cs23}) the SU(3) coherent state structure functions, 
\bea 
F^{\alpha_1..\alpha_{n_1};\beta_1..\beta_{n_2}}\big(z[1],z[2]\big) = 
\frac{1}{n_1!n_2!} z[1]^{\alpha_1} z[1]^{\alpha_2} \ldots z[1]^{\alpha_{n_1}}
z[2]^{\beta_1} z[2]^{\beta_2} \ldots z[2]^{\beta_{n_2}}.
\label{su3sf} 
\eea
are analytic functions of SU(3) group manifold co-ordinates. 
%The SU(3) coherent states (\ref{cs23})  and the structure functions (\ref{su3sf}) are straightforward 
%generalization of  the corresponding SU(2) results (\ref{su2cs2}) and (\ref{su2sf}) respectively. 
Like in SU(2) case, the resolution of identity property follows from the group transformation 
laws. Using the SU(3) transformations (\ref{su3gta}), we find that 
the SU(3) coherent states transform as: 
$|z[1],z[2]\rangle_{[n_1,n_2]} \rightarrow |z^\prime[1],z^\prime[2]\rangle_{[n_1,n_2]}$ 
where  
\bea 
%z^{\alpha}[1] \rightarrow 
{z^{\prime}}^{\alpha}[1] =\left(\exp i (\theta^{\mathrm a} \frac{\lambda^{\mathrm a}}{2})
\right)^{\alpha}{}_{\beta}~z^{\beta}[1], ~~~~~~~~ 
%\nonumber \\  
%z^{\alpha}[2] \rightarrow 
{z^{\prime}}^{\alpha}[2] =\left(\exp i (\theta^{\mathrm a} \frac{\lambda^{\mathrm a}}{2})
\right)^{\alpha}{}_{\beta}~z^{\beta}[2]. 
\label{su3gt} 
\eea
Again like in SU(2) case,  $z[1]~\&~z[2]$ transform like SU(3) triplets, the 
orthonormality conditions (\ref{zc3}) remains invariant under the SU(3) 
transformations. In other words, 
the coherent state (\ref{csc1c2}) defined at a point $(z[1],z[2])$ 
transform to the coherent state at $(z'[1],z'[2])$ on the 
 SU(3) group manifold. 
Therefore, the  operator ${\cal{O}}_{[n_1,n_2]}$: 
\bea
{\mathcal{O}}_{[n_1,n_2]} \equiv
 \int d\mu(z) \left(\big|z[1],z[2]\big\rangle_{[n_1,n_2]}~
{\phantom{\big\rangle}}_{[n_1,n_2]}\big\langle z[1],z[2]\big|\right)
\label{op3} 
\eea
with SU(3) Haar measure $$\int d\mu(z) \equiv \Big(\int d^{2}z[1] d^2z[2]\Big) \Big(
\prod_{\alpha,\beta=1}^2 \delta(z[\alpha].z^{*}[\beta] -\delta_{\alpha,\beta})$$ 
is invariant under all SU(3) transformations (\ref{su3gt}):   
\bea
[Q^{\mathrm a},{\cal{O}}_{[n_1,n_2]}] = 0, ~~~~ \forall {\mathrm a} =1,2,....,8
\eea 
\noindent Therefore, by  Schur's Lemma ${\cal{O}}_{[n_1,n_2]}$ is proportional to identity operator. 

The SU(3) coherent states (\ref{cs23}) and the structure functions (\ref{su3sf}) are straightforward  
generalization of the SU(2) coherent states (\ref{cs2}) and the corresponding structure functions (\ref{su2sf})
respectively.  The latter, in turn, are  SU(2) generalization of the oldest Heisenberg-Weyl or harmonic oscillator 
coherent states (\ref{wcs}) and the associated structure functions (\ref{hwsf}). 
%Thus SU(3) coherent states in terms of SU(3) 
%irreducible Schwinger bosons are as simple to construct as the oldest 
%Heisenberg-Weyl coherent states in terms of harmonic oscillators. 

\section{SU(N) coherent state}
\label{sun} 

Like the previous section on SU(3), this section has two parts. In the first part we briefly 
describe the SU(N) irreducible Schwinger bosons \cite{sunisb} and in the 
second part we exploit it to define SU(N) coherent states.

\subsection{Irreducible Schwinger Boson Representations of SU(N)}

\begin{figure}[t]
\begin{center}
\includegraphics[width=0.7\textwidth,height=0.35\textwidth]
{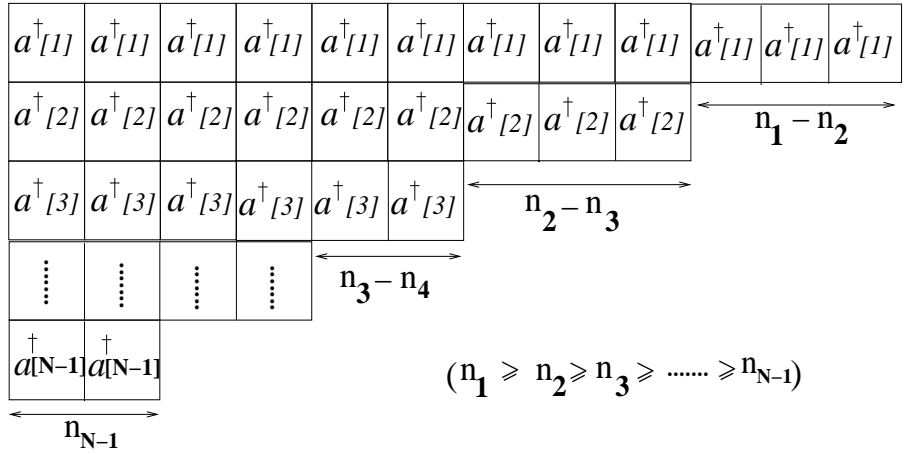}
\end{center}
\caption{SU(N) Young table in the $[n_1,n_2,n_3,\ldots,n_{N-1}]$ representation. The SU(N) irreducible 
Schwinger boson monomial state (\ref{sunir}) and not (\ref{mssun}) carries all the symmetries of this 
SU(N) Young tableau.}   
\label{sunyt}
\end{figure}

The rank of SU(N) group is $(N-1)$. Therefore, the fundamental constituents required to construct any 
arbitrary irrep. of SU(N) can be chosen to be $N-1$ independent Schwinger 
boson $N$-plets: $a^{\dagger\alpha}[1]$, $a^{\dagger\alpha}[2]$, $a^{\dagger\alpha}[3]$,...
,$a^{\dagger\alpha}[N-1]$ with $\alpha =1,2,3,..,N$. The SU(N) generators in terms of these Schwinger 
bosons are:
\bea
\label{sungen}
Q^{\mathrm a}= \sum _{i=1}^{N-1}a^\dagger [i]\,\frac{\Lambda^{\mathrm a}}{2}\,a[i], ~~~~~~~~~~~~
~{\mathrm a} =1,2,\cdots ,(N^2-1). 
\eea
Above $\Lambda^{\mathrm a}$'s are the generalization of Gell-Mann matrices for SU(N). The $N(N-1)$ Harmonic 
oscillators present in (\ref{sungen}) creates a $N(N-1)$ dimensional Hilbert space $\mathcal H^{N(N-1)}_{\mathrm 
{HO}}$. 
The SU(N) transformations are:  
\bea 
a^\dagger_\alpha[i] \rightarrow a^\dagger_\beta[i]~ 
\left(\exp{i \theta^{\mathrm a} {\lambda^{\mathrm a}\over 2}} \right)^\beta_{~~\alpha} ~~~~~~~ i=1,2, \cdots ,(N-1). 
%  a^\alpha[i] \rightarrow \left(exp -i \theta^{\mathrm a} 
%{\lambda^{\mathrm a}\over 2}\right)^\alpha_\beta a^\beta[i] ~~~~~~~~~i =1,2.
\label{sungt}
\eea    
The defining equations (\ref{sungen}) imply that the $(N-1)$ Casimirs associated with SU(N) group, denoted 
by ${\mathcal C}_i$, are the $(N-1)$ number operators 
\bea 
\mathcal N_i\equiv {\cal C}_i= a^{\dagger}[i] \cdot a[i], ~~~~i=1,2,..,(N-1). 
\eea
A particular SU(N) irreducible representation is labeled by their eigenvalues: $\left[n_1,n_2, \ldots  ,n_{N-1}
\right]$. 
\noindent The SU(N) monomial eigenstate: 
\bea 
\Big|\alpha_1^{[1]}.. \alpha_{n_1}^{[1]}; \alpha_1^{[2]} .. \alpha_{n_2}^{[2]}; \cdots ;   
\alpha_1^{[N-1]} .. \alpha_{n_{N-1}}^{[N-1]}\Big\rangle \equiv 
\underbrace{a^\dagger_{\alpha_1^{[1]}}[1] .. a^\dagger_{\alpha_{n_1}^{[1]}}[1]}_{n_1 ~of ~ a^{\dagger}[1] 
.......}      
\cdots \underbrace{a^\dagger_{\alpha_1^{[N-1]}}[N-1] 
..a^\dagger_{\alpha_{n_{N-1}}^{[N-1]}}[N-1]}_{...... n_{N-1} ~ of ~ a^\dagger[N-1]}\big|0\big\rangle,      
\label{mssun} 
\eea 
satisfies   
$${\cal C}_i~ 
\Big|\alpha_1^{[1]}.. \alpha_{n_1}^{[1]}; \alpha_1^{[2]} .. \alpha_{n_2}^{[2]}; \cdots ;   
\alpha_1^{[N-1]} .. \alpha_{n_{N-1}}^{[N-1]}\Big\rangle 
= n_i~
\Big|\alpha_1^{[1]}.. \alpha_{n_1}^{[1]}; \alpha_1^{[2]} .. \alpha_{n_2}^{[2]}; \cdots ;   
\alpha_1^{[N-1]} .. \alpha_{n_{N-1}}^{[N-1]}\Big\rangle.  
$$ 
However, like in SU(3) case (\ref{mssu3}), these states are  SU(N) reducible due to the presence of 
SU(N) invariants. These invariants can be removed by implementing the symmetries of the associated SU(N)   
Young tableau in Figure \ref{sunyt} which contains  $n_i$ boxes in the $i^{th}$ row. 

\noindent Like in SU(3) case we impose the following $(N-1)(N-2)\over2$  constraints \cite{mosh,sunisb}: 
\bea 
\hat L_{ij} =a^\dagger[i]\cdot a[j]\approx 0,~~~~~  i,j = 1,2, \cdots (N-1)~ {\textrm and}~ i < j. 
\label{sunc} 
\eea 
on the monomial states (\ref{mssun}). Like in SU(3) case,  we can  trivialize all the 
constraints in (\ref{sunc}) in terms of SU(N) irreducible Schwinger bosons \cite{sunisb}. 
The SU(N) irreducible Schwinger boson creation operators acting on the vacuum directly create 
states having all the symmetries of the corresponding Young tableaux. 
These new irreducible Schwinger bosons $A^\dagger[i]$ for $i=1,2,..,N-1$ are related to the ordinary Schwinger 
bosons in (\ref{sungen}) as follows: 
\bea
A^{\dagger\alpha}[k] &=& a^{\dagger\alpha}[k]+\sum_{r=1}^{k-1}
\sum_{\{i_1,..,i_r\}=1}^{{k-1}_{\prime}}  ~F^k_{i_1}~F^k_{i_2} \cdots F^k_{i_r} \hat L_{ki_1}~ \hat L_{i_1i_2}
\ldots\hat L_{i_{r-1}i_r}a^{\dagger\alpha}[i_r]. 
\label{sunirsb} 
\eea
In (\ref{sunirsb}) $k=1,2, \cdots (N-1)$ and the prime over the second summation ($\sum^{\prime}$) implies 
that the  ordering $k>i_1>i_2>...>i_r$ has to be maintained.  
The general form of $F^k_i(n_1,..,n_{N-1})$ is given by,
\bea
F^k_i=-\frac{1}{n_i-n_k+1+k-i}. 
\eea
The defining equations (\ref{sunirsb}) imply that the  SU(N) irreducible Schwinger bosons  
also transform as SU(N) N-plets: 
\bea 
A^\dagger_\alpha[i] \rightarrow A^\dagger_\beta[i]~ 
\left(\exp{i \theta^{\mathrm a} {\lambda^{\mathrm a}\over 2}} \right)^\beta_{~~\alpha} ~~~~~~~ i=1,2, \cdots ,(N-1). 
%  a^\alpha[i] \rightarrow \left(exp -i \theta^{\mathrm a} 
%{\lambda^{\mathrm a}\over 2}\right)^\alpha_\beta a^\beta[i] ~~~~~~~~~i =1,2.
\label{sungta}
\eea    
Therefore, as in SU(2) case, the Hilbert space created by the monomials of SU(N) irreducible Schwinger 
boson (\ref{sunirsb}) creation operators is isomorphic to the space of irreducible representations of 
SU(N). In other words the  state   
%\bea 
%A^\dagger[i]\cdot A[j]\approx 0,~~~~\forall~~ i\ne j.
%\label{alz}
%\eea 
%Now, any SU(N) irreducible tensor becomes as simple as that of SU(2) and is obtained as:
\bea
\label{sunir} 
&&
\bigg|\alpha^{[1]}_1,\alpha^{[1]} _2,\ldots,\alpha^{[1]}_{n_1};
\alpha^{[2]}_1,\alpha^{[2]}_2, \ldots ,\alpha^{[2]}_{n_2}; \cdots 
,\alpha^{[N-1]}_1,\alpha^{[N-1]}_2\ldots,\alpha^{[N-1]} _{n_{N-1}}\Bigg\rangle_{[n_1,n_2,\cdots n_{N-1}]}  
 \\  
&\equiv & 
%\underbrace
\underbrace{\left(A^{\dagger \alpha^{[N-1]}_1}[N-1] \cdots A^{\dagger \alpha^{[N-1]}_{n_{N-1}}}[N-1]
\right)}_{n_{N-1} ~ of A^\dagger[N-1]} 
\cdots \cdots  
\underbrace{\left(A^{\dagger \alpha^{[2]}_1}[2] \cdots A^{\dagger \alpha^{[2]}
_{n_2}}[2]\right)}_{n_2 ~ of A^\dagger[2]} 
\underbrace{\left(A^{\dagger \alpha^{[1]}_1}[1] \cdots A^{\dagger \alpha^{[1]}_{n_1}}[1]\right)}_{n_1 ~ of A^\dagger[1]} 
\big|0\big\rangle  \hspace{1.5cm}  \nonumber 
\eea 
carries all the symmetries of an SU(N) Young tableaux shown in Figure \ref{sunyt}.

\subsection{Construction of SU(N) Coherent States}

Like in SU(2) and SU(3) cases in (\ref{su2cons}) and (\ref{zc3}) respectively, we characterize the SU(N) group 
manifold by $N-1$ number of complex $N$-plets:  $\{z_\alpha[i]\}$, $i=1,2,\ldots, N-1$ and  
$\alpha=1,2,\ldots N$ following orthonormality constraints: 
\bea
\label{zc}
\bar z[\alpha]\cdot z[\beta]=\delta_{\alpha,\beta}.
\eea 
With the above  parametrization any SU(N) matrix has the following form: 
\bea
\mathcal U_N = \left( \begin{array}{cccccc} 
z_1[1] & z_1[2] &\ldots & \ldots &  z_1[N-1] & (\bar z[1]\wedge \bar z[2]\wedge\ldots\wedge \bar z[N-1])_1  \\
z_2[1] & z_2[2] &\ldots &\ldots &  z_2[N-1] &  (\bar z[1]\wedge \bar z[2]\wedge\ldots\wedge \bar z[N-1])_2  \\
\vdots &\vdots & \ddots &{ }& \vdots & \vdots \\
\vdots &\vdots & {}&\ddots & \vdots & \vdots \\
z_N[1] & z_N[2] &\ldots &\ldots &  z_N[N-1] &  (\bar z[1]\wedge \bar z[2]\wedge\ldots\wedge \bar z[N-1])_N  \\
\end{array} \right)
\eea

\noindent At this stage we generalize (\ref{su2cs2}) and (\ref{cs23}) to define the SU(N) coherent state 
generating function as: 
\bea 
\big|z[1],z[2],\ldots z[N-1]\big\rangle \equiv \exp\left( z[N-1]\cdot A^{\dagger}[N-1]\right)\ldots\ldots\exp \left(
 z[2]\cdot  A^{\dagger}[2] \right){\exp\left( z[1]\cdot  A^{\dagger}[1]\right) \big|0\big\rangle}.
\label{cs33}
\eea 
Note that the  coherent state generating function (\ref{cs33}) contains all possible irreducible representations 
of SU(N). Further, the expressions for SU(N+1) and SU(N) coherent states differ only by the last exponential 
factor  in (\ref{cs33}). Therefore, the present SU(N) coherent state construction is iterative in nature.  
Now, projecting out a specific coherent state denoted by the  set of particular values of the SU(N) 
Casimirs, i.e., $A^{\dagger}[i]\cdot A[i]$ having eigenvalue $n_i$ with $i=1,2,\cdots (N-1)$ and 
$n_1\ge n_2\ge\ldots\ge n_{N-1}$ we get the SU(N) coherent state in the irreducible representation 
$[n_1,n_2,\cdots,n_{N-1}]$: 
\bea 
\label{csc1c2}
|z[1],z[2],\ldots z[N-1]\rangle_{[n_1,n_2\cdots n_{N-1}]}  \equiv  
\frac{\left( z[N-1]\cdot A^{\dagger}[N-1]\right)^{n_{N-1}}}{n_{N-1}!}\ldots \frac{\left( z[2]\cdot 
A^{\dagger}[2] \right)^{n_2}}{n_2!} \frac{\left( z[1]\cdot  A^{\dagger}[1]\right)^{n_1}}{n_1!} \big|0\big\rangle 
~~~~~~~~~~\\ 
= \sum_{{\alpha}^{[1]}_1,..,{\alpha}^{[1]}_{n_1} =1}^{N} \sum_{{\alpha}^{[2]}_1,..,{\alpha}^{[2]}_{n_2} =1}^{N} 
\sum_{{\alpha}^{[N-1]}_1,..,{\alpha}^{[N-1]}_{n_{N-1}} =1}^{N} 
F^{\alpha^{[1]}_1..\alpha^{[1]}_{n_1} \cdots \alpha^{[N-1]}_1..\alpha^{[N-1]}_{n_{N-1}} }\big(z[1],z[2] 
\cdots z[N-1]\big) ~
\underbrace{\bigg|\alpha^{[1]}_1..\alpha^{[1]}_{n_1} \cdots \alpha^{[N-1]}_1 ..\alpha^{[N-1]}_{n_{N-1}}
\bigg\rangle}_{{\textrm SU(N)~irrep. ~state~(\ref{sunir})~}}{}_{{}_{[n_1,n_1.n_{N-1}]}} \nonumber  
\eea 
In (\ref{csc1c2}) the SU(N) coherent state structure functions are given by: 
\bea 
\label{sfsun}
F^{\alpha^{[1]}_1..\alpha^{[1]}_{n_1} \cdots \alpha^{[N-1]}_1..\alpha^{[N-1]}_{n_{N-1}} }&=& \frac{1}{n_1!n_2!\ldots n_{N-1}!} z[1]^{\alpha^{[1]}_1}...~z[1]^{\alpha^{[1]}_{n_1}}\cdots\cdots z[N-1]^{\alpha^{[N-1]}_1}...~~ z[N-1]^{\alpha^{[N-1]}_{n_{N-1}}}.
\eea
The states in (\ref{csc1c2}) depend smoothly on the SU(N) group manifold coordinates. 
We now check the resolution of identity. 
Like in the previous SU(2) and SU(3) sections, under SU(N) transformations (\ref{sungta}) all the $(N-1)$ 
coherent state co-ordinates  $z[i]$ transform as N-plets: 
% in (\ref{cs33}) or (\ref{csc1c2}) 
\bea
\label{suntrans}
z_\alpha[i]\rightarrow z'_\alpha[i]= z_\beta[i]\left( \exp i \sum_{\mathrm a=1}^{N^2-1} 
\theta^{\mathrm a}\Lambda^{\mathrm a} \right)^\beta _{~~~\alpha}. 
\eea 
We again define the operator ${\mathcal{O}}_{[n_1,n_2,\cdots n_{N-1}]}$ as: 
\bea
{\mathcal{O}}_{[n_1,n_2,\cdots ,n_{N-1}]} \equiv 
\int d\mu(z)\, \left(\big|z[1],z[2],\ldots z[N-1]\big\rangle_{[n_1,n_2,\cdots n_{N-1}]}~
{\phantom{\big\rangle}}_{[n_1,n_2,\cdots n_{N-1}]}\big\langle z[1],z[2],\ldots z[N-1]\big|\right). 
\label{op} 
\eea
In (\ref{op}) $\int d\mu (z)$ is  the SU(N) invariant Haar measure: 
$$\int d\mu(z) \equiv \left[\prod_{\alpha=1}^{N-1} \int d^{2}z[\alpha]\right] 
\prod_{\alpha,\beta} \delta\left(z[\alpha].z^{*}[\beta] -\delta_{\alpha,\beta}\right).$$  
\noindent Under SU(N) transformations (\ref{suntrans}), ${\cal{O}}_{[N]}$ 
remains invariant. Therefore,
%The transformation laws (\ref{suntrans}) imply that the operator: 
\bea
[Q^{\mathrm a},{\cal{O}}_{[n_1,n_2,\cdots ,n_{N-1}]}] = 0, ~~~~ \forall {\mathrm a} =1,2,....,N^2-1. 
\eea 
\noindent The Schur's Lemma implies: 
\bea 
{\cal{O}}_{[n_1,n_2,\cdots, n_{N-1}]} = I_{[n_1,n_2,\cdots ,n_{N-1}]}.
\label{idd}
\eea 
\noindent In (\ref{idd})  $I_{[n_1,n_2,\cdots ,n_{N-1}]}$ is proportional to  identity operator 
in the irreducible representation subspace.  
We again emphasize that the SU(N) coherent states in (\ref{csc1c2}) are the most straightforward extension of 
the Heisenberg Weyl, SU(2) and SU(3) coherent states in (\ref{wcs}), (\ref{su2cs2}) and (\ref{cs23}) 
respectively.

\section{Conclusions}

We have exploited SU(N) irreducible Schwinger boson creation operators to construct SU(N) coherent states. This 
construction is analogous to the simplest and the oldest harmonic oscillator coherent state construction.  
This procedure is iterative in N. It is also self contained as it does not require any prior knowledge of SU(N) 
group elements and their representations. This novel SU(N) coherent state construction can be used to compute 
SU(N) Clebsch-Gordan and recoupling coefficients. This amounts to generalizing the Schwinger method to compute 
these coefficients for SU(2) (see section 3 of \cite{sch} on the addition of angular momenta) to SU(N). 
This is particularly interesting as these SU(N) coupling coefficients for arbitrary N are not yet known in 
closed form.  The work in this direction is in progress and will be reported elsewhere.

\end{document}